\documentclass[prb,aps,showpacs,floatfix,floats,nobalancelastpage,twocolumn]{revtex4-1}
\usepackage{graphicx}
\usepackage{graphics}
\usepackage{latexsym,amsmath,amssymb,bm,euscript}
\usepackage{color}
\usepackage{dcolumn}
\usepackage{bm}
\usepackage{epstopdf}
\usepackage{times}
\usepackage{multirow}
\usepackage{wasysym}
\usepackage{subfigure}
\def\etal{~\textit{et~al.}}
\def\ra{\rangle}
\def\la{\langle}

\def\Hc{{\rm H.c.}}

\def\ET{{$\kappa$-(ET)$_2$Cu$_2$(CN)$_3$}}
\def\dmit{{EtMe$_3$Sb[Pd(dmit)$_2$]$_2$}}

\begin{document}

\title{SU(2)-invariant Majorana spin liquid with stable parton Fermi surfaces in an exactly solvable model }
\author{Hsin-Hua Lai}
\affiliation{Department of Physics, California Institute of Technology, Pasadena, California 91125, USA}
\author{Olexei I. Motrunich}
\affiliation{Department of Physics, California Institute of Technology, Pasadena, California 91125, USA}
\date{\today}

\begin{abstract}
We construct an exactly solvable spin-orbital model on a decorated square lattice that realizes an SU(2)-invariant Majorana spin liquid with parton Fermi surfaces, of the kind discussed recently by Biswas\etal [ Phys. Rev. B. {\bf 83}, 245131(2011)].  We find power-law spin correlations as well as power-law spin-nematic correlations with the same dominant $1/|{\bf r}|^3$ envelope.  The model is solvable also in the presence of Zeeman magnetic field.  One fermion species carries $S^z=0$ quantum number and its Fermi surface is not altered in the field, while the Fermi surfaces of the other species evolve and can disappear.  In particular, we find an interesting half-magnetization plateau phase in which spin excitations are gapful while there remain spinless gapless excitations that still produce metal-like thermal properties.  In the fully-magnetized phase, the model reduces to the one proposed by Baskaran\etal, [arXiv:0908.1614v3] in terms of the orbital degrees of freedom. 
\end{abstract}
\maketitle

\section{Introduction}
Gapless quantum spin liquids (QSL)\cite{LeeNagaosaWen, Polchinski94, Altshuler94, YBKim94, SSLee2009, Metlitski2010, Mross2010, DBL, Rantner2002, Hermele2004, SenthilFisher_Z2, PWA_science, BZA, Ioffe1989, Lee08_science, Balents_nature} are perhaps some of the most intriguing fractionalized phases.  Much interest in these is motivated by recent experimental realizations in two-dimensional (2D) organic compounds \dmit~and \ET.\cite{Shimizu03, Kurosaki05, SYamashita08, MYamashita09, McKenzie, Itou07, Itou08, Itou10, MYamashita10, Powell10, SYamashita11,Kanoda11_report}  One proposal with Gutzwiller-projected Fermi sea wave function\cite{ringxch, LeeandLee05} is an appealing candidate but does not appear to be able to capture all experimental phenomenology. Searching for alternatives, many possible proposals have been presented.\cite{Grover09, Qi09, Xu09, Biswas11} Very recently, Biswas\etal\cite{Biswas11} proposed an SU(2)-invariant Majorana QSL, which we find fascinating and in need of more attention. Motivated by this proposal, here we want to realize such long-wavelength QSL in an exactly solvable microscopic model. Following the route discovered by Kitaev\cite{Kitaev06} and generalized to produce many other exactly solvable models,\cite{Wen03, Feng07, Baskaran07, Lee07, Yao07, Chen07, Vidal08, Yao09, Mandal09, Wu09, Nussinov09, Baskaran09, Willans10, Tikhonov10, Tikhonov10-arxiv, Chern10, Wang10, Chua10, Yao10, Dhochak10} in particular with SU(2) spin invariance\cite{Wang10, Yao10} or with parton Fermi surfaces,\cite{Baskaran09, Tikhonov10, Chua10} we find a Kitaev-type model with both SU(2)-invariance and parton Fermi surfaces.

Our model is realized using both spin-1/2 and orbital degrees of freedom\cite{Fawang09,Yao10} at each site of a decorated square lattice.\cite{Baskaran09}  The system can be reduced to three species of free Majorana fermions coupled to background $Z_2$ gauge fields such that it is exactly solvable and parton Fermi surfaces are realized. We formulate long wavelength description in terms of an occupied Fermi pocket of three complex fermions ($f^x,~f^y,~f^z$) that transform as a vector under spin rotation. For general illustration (and also for preventing possible pairing instabilities away from the exactly solvable limit), we consider a model that lacks time-reversal and lattice inversion symmetries.  Because of the exact solvability, we can learn much reliable physics information about such Majorana QSL.

Specifically, we study spin correlations and spin-nematic correlations in our model.  The main result is that these correlations have the {\it same} dominant power-law behaviors with $1/|{\bf r}|^3$ envelope in real space and oscillations at incommensurate wavevectors which form what we call singular surfaces\cite{LeeNagaosaWen, Altshuler94, DBL} in the momentum space. Because of the $Z_2$ nature of the QSL and the absence of the time-reversal and inversion symmetries, there are additional non-trivial $\pm ({\bf k}_{FR}+{\bf k}_{FL})$ and $\pm 2 {\bf k}_{F}$ critical surfaces besides the more familiar ${\bf k}_{FR}-{\bf k}_{FL}$ surface in the correlations.

The model is still exactly solvable in the presence of Zeeman magnetic field.  An interesting property is that the Zeeman field only couples to the $f^x$ and $f^y$ fermions while the $f^z$ fermion remains unaltered and therefore the $f^z$ Fermi surface remains and always gives gapless excitations.  We calculate the magnetization as a function of magnetic field.  Interestingly, there is a plateau phase in which the spins are half-polarized with short-ranged spin correlations while the Fermi surface of $f^z$ still exists and gives gapless excitations, which can be detected using local energy operator like bond energy.\cite{Lai11_bondcorr}

The paper is organized as follows. In Sec.~\ref{Sec:SU(2)_case}, we define the model on the decorated square lattice and solve it and discuss qualitative properties of the spin liquid phase. In Sec.~\ref{Sec:Correlations}, we define the spin correlation functions and spin-nematic correlation functions. In Sec.~\ref{Subsec:long_wl} we provide a theoretical approach to describe the long-distance behavior of the correlations. In Sec.~\ref{Subsec:Data4SU(2)}, we present exact numerical calculations of the spin correlations and spin-nematic correlations. In Sec.~\ref{Sec:Zeeman}, we consider our model in the presence of the Zeeman magnetic field and specifically calculate the magnetization curve as a function of the field. We conclude with some discussion.

\section{SU(2)-invariant Majorana spin liquid with stable Fermi surfaces}\label{Sec:SU(2)_case}
\begin{figure}[t]
\includegraphics[width=\columnwidth]{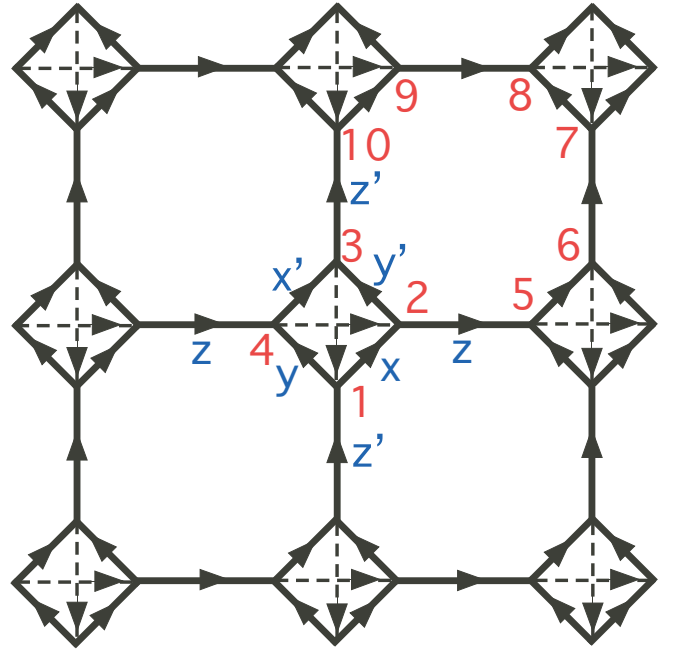}
\caption{Graphical representation of the exactly solvable Kitaev-type model and its solution in the zero flux sector.  The $c^{x,y,z}$ Majoranas propagate with pure imaginary hopping amplitudes specified by the couplings $J^x,~J^y,~J^z,~J^{x'},~J^{y'},$ and $J^{z'}$; the signs in our chosen gauge are indicated by the arrows.
}
\label{graphical_reps}
\end{figure}

Motivated by the ideas from Baskaran\etal\cite{Baskaran09}, Yao\etal\cite{Yao10}, and Wang\cite{Wang10}, we construct an exactly solvable Kitaev-type model including both orbital and spin degrees of freedom with spin-rotation invariance.  The Hamiltonian is
\begin{eqnarray}\label{Def:modelH}
\mathcal{H}=\mathcal{H}_{0}+H_{TRB}+K_{\Diamond} \sum_{\Diamond} W_{\Diamond} +K_{\octagon} \sum_{\octagon} W_{\octagon}~,
\end{eqnarray}
where,
\begin{eqnarray}
&& \mathcal{H}_{0}=\sum_{\lambda-{\rm link}~\langle j k \rangle} J^{\lambda}_{jk} \left(\tau^\lambda_j \tau^{\lambda}_k \right) \left( \vec{\sigma}_j \cdot \vec{\sigma}_k \right),~\\
\nonumber && \mathcal{H}_{TRB}=\frac{h}{2}\sum_{\Diamond} \bigg{[} \left(\tau^{x}_3 \tau^{z}_4 \tau^y_1-\tau^{x}_1 \tau^{z}_2 \tau^{y}_3\right) \left( \vec{\sigma}_3 \cdot \vec{\sigma}_1\right)~\\
&& \hspace{2.5cm}+\left(\tau^y_4 \tau^z_1 \tau^x_2 - \tau^y_2\tau^z_3\tau^x_4\right)\left( \vec{\sigma}_{4}\cdot\vec{\sigma}_{2}\right)\bigg{]},~\label{HTRB}\\
&& W_{\Diamond} = \tau^{z}_{1}\tau^{z}_{2}\tau^{z}_{3}\tau^{z}_{4},~\label{diamondp}\\
&& W_{\octagon}=\tau^{x}_{3}\tau^{x}_{2}\tau^{y}_{5}\tau^{y}_{6}\tau^{x}_{7}\tau^{x}_{8}\tau^{y}_{9}\tau^{y}_{10}.~\label{octagonp}
\end{eqnarray}
The graphical representation of the model is shown in Fig.~\ref{graphical_reps}.  At each site of the decorated square lattice, there are spin and orbital degrees of freedom.  $\mathcal{H}_0$ is a Kugel-Khomskii-like Hamiltonian with $\vec{\sigma}$ being the spin-1/2 Pauli matrices and $\vec{\tau}$ being the Pauli matrices acting on the orbital states.\cite{Fawang09, Yao10}  The site labels in Eqs.~(\ref{HTRB})-(\ref{octagonp}) are shown in Fig.~\ref{graphical_reps}.  $\mathcal{H}_{TRB}$ represents an additional Time-Reversal-Breaking (TRB) interaction in the small diamonds\cite{Yao10} (in principle, all four terms in the square brackets can have independent couplings).  The reason for introducing the TRB and allowing different $J^\lambda$ couplings in $\mathcal{H}_0$ that break the lattice point group symmetries is to avoid worrying about Cooper pair instabilities of the parton Fermi surface away from the exactly solvable limit.

In addition, there are two types of elementary plaquettes (square and octagon) in the decorated square lattice (Fig.~\ref{graphical_reps}), and two types of local conserved operators, $W_{\Diamond}$ for the squares and $W_{\octagon}$ for the octagons in Eq.~(\ref{diamondp}) and Eq.~(\ref{octagonp}).  The plaquette operators $W_p$ commute among themselves and with all other terms in the Hamiltonian and the $K_p$ terms are added to stabilize particular flux sector (see Fig.~\ref{graphical_reps}).

Introducing Majorana representation of spin-1/2,\cite{Tsvelik92, Coleman94, Shastry97} we write the spin and orbital operators as
\begin{eqnarray}
\sigma^{\alpha}_j=-\frac{i}{2}\sum_{\beta,\gamma}\epsilon ^{\alpha \beta \gamma} c^\beta_j c^\gamma_j,~\label{Majorana_reps:spin}\\
\tau^{\alpha}_j=-\frac{i}{2}\sum_{\beta,\gamma}\epsilon ^{\alpha \beta \gamma} d^\beta_j d^\gamma_j.~\label{Majorana_reps:orbital}
\end{eqnarray}

On each site $j$ of the decorated square lattice, we realize the physical four-dimensional Hilbert space using six Majorana fermions $c^{x}_j$, $c^{y}_j$, $c^{z}_j$, $d^{x}_j$, $d^{y}_j$, and $d^{z}_j$, with the constraint $D_j \equiv -i c^x_j c^y_j c^z_j d^x_j d^y_j d^z_j =1$ (namely, for any physical state $|\Phi \ra_{\rm phys}$, we require $D_j |\Phi \ra_{\rm phys} = |\Phi \ra_{\rm phys}$). Therefore, $\sigma_j^\alpha \tau_j^\beta |\Phi \ra_{\rm phys} = i c_j^\alpha d_j^\beta  |\Phi \ra_{\rm phys}$.

In terms of the Majoranas, the Hamiltonian can be rephrased as
\begin{eqnarray}
&& \mathcal{H}_0 = i\sum_{\langle j k \rangle} \hat{u}_{jk} J_{jk} \sum_{\alpha=x,y,z}c^{\alpha}_{j}c^{\alpha}_{k},\\
&& \mathcal{H}_{TRB} = i \frac{h}{2} \sum_{\Diamond} \bigg{[} \left(\hat{u}_{34} \hat{u}_{41} + \hat{u}_{12} \hat{u}_{23} \right) \sum_{\alpha=x,y,z} c^\alpha_3 c^\alpha_1 \\
&& \hspace{2.4cm} - \left(\hat{u}_{41} \hat{u}_{12} + \hat{u}_{23} \hat{u}_{34} \right) \sum_{\alpha=x,y,z} c^\alpha_4 c^\alpha_2 \bigg{]},~~~~~\\
&& W_{p=\{\Diamond,\octagon\}}=-\prod_{\la jk \ra \in p}\hat{u}_{jk},~
\end{eqnarray}
where $\hat{u}_{jk} \equiv -id^{\lambda}_j d^{\lambda}_k$ for $\lambda$-link $\la j k \ra$. Following familiar analysis in Kitaev-type models, we observe that in the enlarged Hilbert space, $\hat{u}_{jk}$ commute among themselves and with the Hamiltonian, and we can proceed by replacing them by their eigenvalues $\pm 1$ and interpreting as static $Z_2$ gauge fields.  The $W_{p}$ terms, with $K_p>0$ and assumed to be sufficiently large, can be used to stabilize the sector with zero fluxes through all elementary plackets, and this can produce parton Fermi surfaces.\cite{Baskaran09}  
In our work, we fix the gauge by taking $u_{jk}=1$ for bonds $j\rightarrow k$ as shown by the arrows in Fig.~\ref{graphical_reps}.  There are four physical sites per unit cell, so for each species $c^\alpha$, $\alpha=x, y, z$, there are four Majoranas per unit cell.  From now on, we replace the site labeling $j$ with $j=\{ {\bf r},a\}$, where ${\bf r}$ runs over the Bravais lattice of unit cells of the decorated square network and $a$ runs over the four sites in the unit cell.  The Hamiltonian can be written in a concise form,
\begin{eqnarray}
\nonumber \mathcal{H}&=&\sum_{\alpha}\sum_{\la jk \ra}c^{\alpha}_{j}\mathcal{A}_{jk}c^{\alpha}_{k}~\\
&=&\sum_{\alpha}\sum_{\la ({\bf r},a),({\bf r'},a')\ra} c^{\alpha}_{{\bf r},a}\mathcal{A}_{{\bf r},a;{\bf r'},a'}c^\alpha_{{\bf r'},a'}.~\label{Kitaev-type H}
\end{eqnarray}
There is translational symmetry between different unit cells, and $\mathcal{A}_{{\bf r},a;{\bf r'},a'}=\mathcal{A}_{aa'}({\bf r}-{\bf r'})$.

In order to give a concise long-wavelength description, it will be convenient to use familiar complex fermion fields.  To this end, we can proceed as follows.  For a general Majorana problem specified by an antisymmetric pure imaginary matrix $\mathcal{A}_{jk}$, we diagonalize $\mathcal{A}_{jk}$ for spectra, but only half of the bands are needed while the rest of the bands can be obtained by a specific relation and are redundant.  Explicitly, for a system with $2m$ bands, we can divide them into two groups.  The first group contains bands from $1$ to $m$ with eigenvector-eigenenergy pairs $\{\vec{v}_{b,{\bm k}}, \epsilon_{b,{\bm k}}\}$, where $b = 1, 2, \dots, m$ are band indices, and the second group contains bands from $m+1$ to $2m$ related to the first group, $\{\vec{v}_{b'=m+b, {\bm k}}, \epsilon_{b'=m+b, {\bm k}}\} = \{ \vec{v}^*_{b, -{\bm k}}, -\epsilon_{b, -{\bm k}}\}$. Using only the bands with $b=1$ to $m$, we can write the original Majoranas in terms of usual complex fermions as
\begin{eqnarray}\label{usual_fermion}
c^{\alpha}({\bf r},a) = \sqrt{\frac{2}{N_{uc}}}\sum_{b=1}^m \sum_{{\bf k} \in {\bf B.Z.}}\left[ e^{i {\bf k}\cdot {\bf r}}v_{b,{\bf k}}(a)f^{\alpha}_{b} ({\bf k}) +\Hc \right],~~~~~~
\end{eqnarray}
where $N_{uc}$ is the number of unit cells, and the complex fermion field $f$ satisfies the usual anticommutation relation, $\{ f^{\alpha\dagger}_{b}({\bf k}),f^{\alpha'}_{b'}({\bf k'})\}=\delta_{\alpha \alpha'}\delta_{b b'}\delta_{{\bf k}{\bf k'}}$. Note that in this SU(2)-invariant model, the eigenvectors for each spin species are the same, $v^{\alpha}_{b,{\bf k}}=v_{b,{\bf k}}$.  In terms of the complex fermion fields, the Hamiltonian becomes
\begin{eqnarray}\label{usual_fermion_H}
\mathcal{H}=\sum_{b=1}^{m}\sum_{{\bf k}\in {\bf B.Z.}} 2\epsilon_{b}({\bf k}) \left[ f^{\alpha \dagger}_{b}({\bf k})f^{\alpha}_b ({\bf k})-\frac{1}{2}\right].
\end{eqnarray}

In the present case, $2m = 4$ and therefore two bands are sufficient to give us the full solution of the Majorana problem. Depending on the parameters, the model can realize different gapped and gapless phases. The latter generally have Fermi surfaces, and here we are focusing on such gapless phases and their qualitative properties. For all illustrations below, we use parameters $\{ J_x,~J_y,~J_z,~J'_x,~J'_y,~J'_z,~h \} = \{1.7,~1.4,~0.4,~1.0,~1.3,~0.2,~0.95 \}$ with $x$, $y$, $z$, $x'$, $y'$, and $z'$ defined in Fig.~\ref{graphical_reps}.  Gapless phases with Fermi surfaces appear in wide parameter regimes, and we remark that there is no fine tuning of parameters to find such phases. The reason we choose to present the specific parameters is that in this case, the Fermi surfaces are sufficiently small, so when we analyze the singularities in the structure factors in Sec.~\ref{Subsec:Data4SU(2)}, it is easier to clearly see the locations of the singularities.

\begin{figure}[t]
\subfigure[ Bands of complex fermions, $f^{x,y,z}_b({\bf k})$, along a cut at $k_y=-3 \pi/4$.]{\label{bands spectrums} \includegraphics[width=\columnwidth]{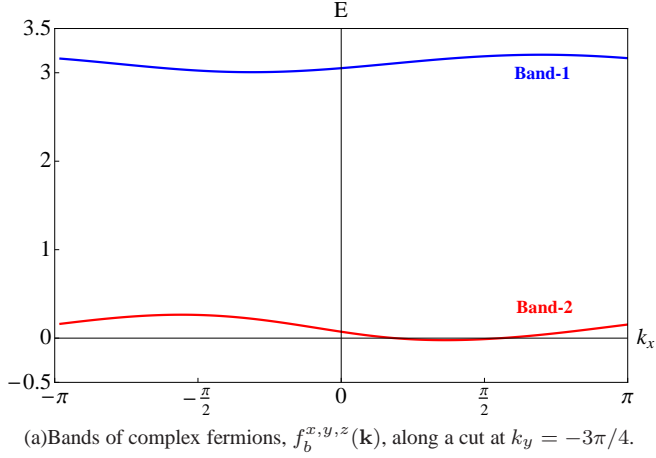}}
\subfigure[ Contour plot of band 2 of complex fermions, $f^{x,y,z}_2({\bf k})$.]{\label{fermi_surface}\includegraphics[width=\columnwidth]{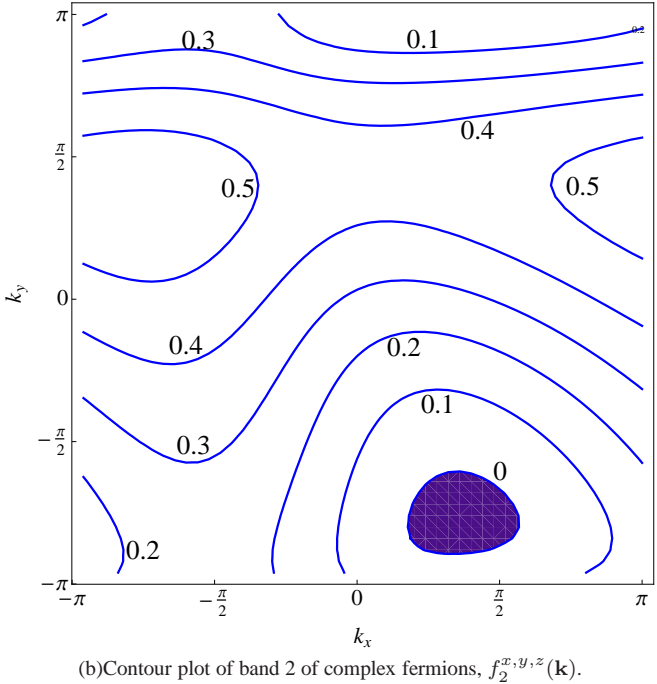}}
\caption{(a) Illustration of energy spectra of the two bands of the complex fermion fields along a cut with $k_y = -3 \pi/4$.  Here we take parameters $\{ J_x, J_y, J_z, J'_x, J'_y, J'_z, h \} = \{1.7 , 1.4, 0.4, 1.0, 1.3,0.2, 0.95 \}$.  Diagonalizing the $\mathcal{A}_{ij}$ matrix in Eq.~(\ref{Kitaev-type H}) gives four bands, but only the two bands shown are needed for solving the Majorana problem (see text).  Because only band 2 crosses zero energy, we simply focus on it for long wavelength analysis. (b) Contour plot of band 2, with the occupied Fermi pocket shaded.
}  
\label{Complex_fermion_bands}
\end{figure}

For an illustration of how these two bands of usual complex fermion fields vary with momentum ${\bm k}$, we show them in Fig.~\ref{bands spectrums} along a cut with $k_y= -3 \pi/4 $.  We label the bands from top to bottom as 1 to 2. We can see that only the band 2 crosses the zero energy, which is true also when we scan the whole $k_y$ axis, and the populated Fermi pocket in the {\bf B.Z.} is shown shaded in Fig.~\ref{fermi_surface}.\cite{footnote_edge}

It is interesting to discuss qualitatively the thermodynamic properties in this phase.  Because of the presence of the gapless Fermi surface, such spin liquid is expected to show metal-like specific heat and spin susceptibility at low temperature, although the Wilson ratio is different from that of spin-1/2 fermions.\cite{Biswas11}
Furthermore, magnetic impurities coupled to this model would possibly show an unusual Kondo effect and Ruderman-Kittel-Kasuya-Yosida interaction.\cite{Dhochak10}

\section{Correlation functions}\label{Sec:Correlations}
We mainly focus on the spin correlations and spin-nematic correlations, and since there are four sites per unit cell, there are many correlation functions one can define. However, since all the spin correlations show similar behaviors among themselves and so do spin-nematic correlations, we consider specific examples defined as
\begin{eqnarray}
&& \mathcal{F}_{1}({\bf r}) \equiv \la S^{+} ({\bf r},2) S^{-} ({\bf 0},2) \ra,~\label{spin0}\\
&& \mathcal{F}_{2}({\bf r}) \equiv \la P^{+} ({\bf r}) P^{-}(0) \ra,~ \label{spin2}
\end{eqnarray}
with 
\begin{eqnarray}
S^{+/-}\equiv S^{x}\pm i S^{y}=(\sigma^{x}\pm i\sigma^{y})/2,
\end{eqnarray}
 and magnon-pair creation operator
\begin{eqnarray}\label{def:spin-nematic}
\nonumber P^{+}({\bf r}) &\equiv& S^{+} ({\bf r},2) \; S^{+} ({\bf r}+\hat{x},4)\\~
&=&S^{+}({\bf r}_{c}-\frac{ {\bm \xi}}{2},2) \; S^{+}({\bf r}_{c} +\frac{{\bm \xi}}{2},4),
\end{eqnarray}
associated with the bond $\la {\bf r},2; {\bf r}+\hat{x},4 \ra$.  In the last line, ${\bf r}_{c} \equiv {\bf r}+\hat{x}/2$ is the center of mass coordinate, and ${\bm \xi} $ is the vector joining the two sites of the nematic operator, which is simply ${\bm \xi} = \hat{x}$ here.  The magnon-pair operator, $P^{+}_{jk} \equiv S^{+}_j S^{+}_k$ for a local pair of sites $\{ j,k \}$, describes spin-nematic properties\cite{Chubukov91, Shannon06, Shindou09} and can be connected to the usual traceless rank two quadrupolar tensor defined as
\begin{eqnarray}\label{quadrupolar_tensor}
\mathcal{Q}^{\alpha \beta}_{jk}=\frac{1}{2} \left( S^{\alpha}_j S^{\beta}_k + S^{\beta}_j S^{\alpha}_k \right) - \frac{1}{3} \delta^{\alpha \beta}\la {\bf S}_j \cdot {\bf S}_k \ra,
\end{eqnarray}
through $P^{+}_{jk} = \mathcal{Q}^{xx}_{jk}-\mathcal{Q}^{yy}_{jk}+2i\mathcal{Q}^{xy}_{jk}.$

Furthermore, power-law correlations in real space correspond to singularities in momentum space, which we can study by considering the corresponding structure factors
\begin{eqnarray}\label{structure_factor}
\mathcal{D}_{1/2}({\bf q})\equiv\sum_{{\bf r}} \mathcal{F}_{1/2} ({\bf r}) e^{-i {\bf q}\cdot {\bf r}}.
\end{eqnarray}

\subsection{Long wavelength analysis}\label{Subsec:long_wl}
We focus on the long-distance behavior and therefore retain only the contribution from band-2. The spin operator can be compactly written as 
\begin{eqnarray}
\nonumber && S^{\alpha}({\bf r},a)\simeq~\\
\nonumber && \simeq  \sum_{{\bf k},{\bf k'}\in {\bf B.Z.}}\sum_{\beta,\gamma} \bigg{\{} N_{{\bf k}{\bf k'}} (a) \epsilon ^{\alpha \beta \gamma} f^{\beta\dagger}_{2}({\bf k}) f^{\gamma}_{2 }({\bf k'}) e^{-i({\bf k}-{\bf k'})\cdot {\bf r}}\\
\nonumber &&\hspace{0.8cm} +\bigg{[} \frac{M_{{\bf k}{\bf k'}}(a)}{2}\epsilon^{\alpha \beta \gamma}f^{\beta}_{2}({\bf k}) f^{\gamma}_{2}({\bf k'})e^{i ({\bf k}+{\bf k'}) \cdot {\bf r}}+\Hc \bigg{]} \bigg{\}},~~~~~~
\end{eqnarray}
where $M_{{\bf k k'}}=-iv_{2,{\bf k}}(a) v_{2,{\bf k'}}(a)/N_{uc}$ and $N_{{\bf k k'}}=-i v^{*}_{2, {\bf k}}(a) v_{2, {\bf k'}}(a)/N_{uc}=-N^*_{{\bf k'} {\bf k}}$. 

In order to determine long-distance behavior at separation ${\bm r}$, we focus on patches near the Fermi surface of band 2 where the group velocity is parallel or antiparallel to the observation direction $\hat{\bm n} = {\bm r}/|{\bm r}|$, because at large separation $|{\bm r}| \gg k_F^{-1}$, the main contributions to the correlations come precisely from such patches.  Specifically, we introduce Right(R) and Left(L) Fermi patch fields and the corresponding energies
\begin{eqnarray}\label{fermi patches:momentum space}
&& f^{\alpha,(\hat{\bm n})}_P(\delta {\bm k}) = f^{\alpha}_2({\bm k}^{(\hat{\bm n})}_{FP} + \delta {\bm k}) ~,\\
&& \epsilon^{(\hat{\bm n})}_P (\delta {\bm k}) = |{\bm v}^{(\hat{\bm n})}_{FP}| \left( P \delta k_\parallel + \frac{\mathfrak{C}^{(\hat{\bm n})}_P}{2} \delta k_\perp^2 \right) ~, \label{fermi patches:energy}
\end{eqnarray}
where the superscript $(\hat{\bm n})$ refers to the observation direction and $P = R/L = +/-$; ${\bm v}^{(\hat{\bm n})}_{FP}$ is the corresponding group velocity (parallel to $\hat{\bm n}$ for the Right patch and anti-parallel for the Left patch); $\mathfrak{C}_{P = R/L}$ is the curvature of the Fermi surface at the Right/Left patch; $\delta k_\parallel$ and $\delta k_\perp$ are respectively components of $\delta {\bm k}$ parallel and perpendicular to $\hat{\bm n}$.  It is convenient to define fields in real space
\begin{eqnarray}\label{fermi patches:real space}
f^{\alpha,(\hat{\bm n})}_P({\bm r}) \sim \sum_{\delta {\bm k} \in {\rm Fermi~Patch}} f^{\alpha,(\hat{\bm n})}_P(\delta {\bm k}) e^{i\delta{\bm k} \cdot {\bm r}} ~,
\end{eqnarray}
which vary slowly on the scale of the lattice spacing [and from now on we will drop the superscript $(\hat{\bm n})$].  In this long-wavelength analysis, the relevant terms in the spin operator are
\begin{eqnarray}
\nonumber &&  S^{\alpha}({\bf r},a)\sim~\\
\nonumber &&  \sim \sum_{P,P'}\sum_{\beta,\gamma}\bigg{\{} N_{P P'} (a) \epsilon ^{\alpha \beta \gamma} f^{\beta\dagger}_{P}({\bf r}) f^{\gamma}_{ P'}({\bf r}) e^{-i({\bf k}_{FP}-{\bf k}_{FP'})\cdot {\bf r}}+~~~~~\\
&& +\bigg{[} \frac{M_{PP'}(a)}{2}\epsilon^{\alpha \beta \gamma}f^{\beta}_{P}({\bf r})f^{\gamma}_{P'}({\bf r})e^{i ({\bf k}_{FP}+{\bf k}_{FP'})\cdot {\bf r}}+\Hc \bigg{]} \bigg{\}},~\label{long_wl:spin}
\end{eqnarray}

The above long-wavelength expression for the $S^{\alpha}$ operator implies that the corresponding correlation function defined in Eq.~(\ref{spin0}) contains contributions with ${\bf q}=0$, ${\bf k}_{FR}- {\bf k}_{FL}$, $\pm 2{\bf k}_{F}$, and $\pm({\bf k}_{FR}+{\bf k}_{FL})$. 
More explicitly, for a patch specified by $\epsilon_P (\delta {\bm k})$ in Eqs.~(\ref{fermi patches:momentum space})-(\ref{fermi patches:energy}), we can derive the Green's function for the continuum complex fermion fields as
\begin{eqnarray}\label{long-wavelength:green function}
\la f^{\alpha\dagger}_{R/L}({\bm 0}) f^{\alpha}_{R/L}({\bm r}) \ra = \frac{\exp[\mp i \frac{3\pi}{4}]}{2^{3/2} \pi^{3/2} \mathfrak{C}_{R/L}^{1/2} |{\bm r}|^{3/2}} ~.
\end{eqnarray}
Using this and Eq.~(\ref{long_wl:spin}), we can obtain the spin correlation
\begin{eqnarray}
\mathcal{F}_{1}({\bf r}) &\sim& -\frac{|N_{RR}|^2}{\mathfrak{C}_{R} |{\bf r}|^3} - \frac{|N_{LL}|^2}{\mathfrak{C}_{L}|{\bf r}|^3}~\label{uniform:spin}\\
&+& \frac{2 |N_{RL}|^2  \sin [ ( {\bf k}_{FR}-{\bf k}_{FL})\cdot {\bf r}] }{\mathfrak{C}_{R}^{1/2}\mathfrak{C}_{L}^{1/2} |{\bf r}|^3} ~\label{krminuskl:spin}\\
&-& \frac{|M_{RR}|^2 \sin (2 {\bf k}_{FR}\cdot {\bf r})}{\mathfrak{C}_{R}|{\bf r}|^3}  + \frac{|M_{LL}|^2 \sin(2 {\bf k}_{FL}\cdot {\bf r})}{\mathfrak{C}_{L}|{\bf r}|^3} ~~~~~~~\label{2kf:spin}\\
&+& \frac{  2|M_{RL}|^2 \cos [ ({\bf k}_{FR}+{\bf k}_{FL})\cdot {\bf r}]}{\mathfrak{C}_{R}^{1/2}\mathfrak{C}_{L}^{1/2} |{\bf r}|^3},~\label{krpluskl:spin}
\end{eqnarray}
where we used $N_{LR} = -N_{RL}^*$ and $M_{LR} = M_{RL}$.

For the long-wavelength description of the spin-nematic correlations, we can in principle plug the expression of spin operator, Eq.~(\ref{long_wl:spin}), into either Eq.~(\ref{def:spin-nematic}) or  Eq.~(\ref{quadrupolar_tensor}).  We remark that even though the microscopic spin-nematic operators contain four local Majorana fermions expressed in general as $c^{\alpha}_{j} c^{\beta}_{k} c^{\gamma}_{j} c^{\delta}_{k}$, when calculating the correlation functions, there are cases when pairs of Majorana fermions Wick-contract locally and produce a constant factor.  Take $\mathcal{Q}^{xy}$ as an example, $\mathcal{Q}^{xy}_{jk} \sim -(c^{x}_j c^{y}_k + c^{y}_j c^{x}_k) c^{z}_j c^{z}_k$, and observe that the last two Majoranas $c^z_j c^z_k$ can Wick-contract when calculating the correlation functions.  For this reason, the spin-nematic correlations show the same dominant power-law behavior as spin correlations, and effectively we have fermion bilinear contributions to the spin-nematic.

From now on, we focus on the dominant contributions to the spin-nematic correlations.  The diagonal and off-diagonal elements of the quadrupolar tensor can be written concisely using center of mass and relative coordinates, where we define $(j,k)=(\{{\bf r},a\},\{ {\bf r'},a' \})=(\{ {\bf r}_c - {\bm \xi}/2, a \}, \{ {\bf r}_c +{\bm \xi}/2,a'\})$,
\begin{eqnarray}
\nonumber 
\mathcal{Q}^{\alpha \alpha} &\sim& \sum_{PP'} \sum_{\beta \neq \alpha} \bigg{\{} \bigg{[} A^{aa'}_{PP'} e^{i ({\bf k}_{FP} + {\bf k}_{FP'}) \cdot {\bf r}_c} f^\beta_P f^\beta_{P'} \\
&&\hspace{0.6cm}+B^{aa'}_{PP'} e^{-i ({\bf k}_{FP} - {\bf k}_{FP'}) \cdot {\bf r}_c} f^{\beta\dagger}_P f^\beta_{P'} \bigg{]} + \Hc\bigg{\}}, \label{quadrupolar:dia} \\
\nonumber\mathcal{Q}^{\alpha \beta} &\sim& \frac{1}{2}\sum_{PP'}
\bigg{\{} \bigg{[} A^{aa'}_{PP'}e^{i({\bf k}_{FP}+{\bf k}_{FP'})\cdot {\bf r}_c}\big{(}f^{\alpha}_{P} f^{\beta}_{P'} + \alpha \!\leftrightarrow\! \beta \big{)} \\
+&&\!\!\!\!\!\!\!B^{aa'}_{PP'}e^{-i({\bf k}_{FP}-{\bf k}_{FP'})\cdot {\bf r}_c} \big{(}f^{\alpha \dagger}_{P} f^{\beta}_{P'} + \alpha \!\leftrightarrow\! \beta \big{)} \bigg{]} + \Hc\bigg{\}}.~~~~~~~
\end{eqnarray}
Here, $A^{aa'}_{PP'} \equiv -i v_{2,P}(a) v_{2,P'}(a')e^{-i({\bf k}_{FP}-{\bf k}_{FP'})\cdot {\bm \xi}/2}/N_{uc}$ and $B^{aa'}_{PP'} \equiv - i v^{*}_{2,P}(a)v_{2,P'}(a')e^{i({\bf k}_{FP}+{\bf k}_{FP'})\cdot {\bm \xi}/2}/N_{uc}$; the slowly varying fermion fields are evaluated at ${\bf r}_c$. The above implies that the spin-nematic correlations contain dominant contributions at wavevectors ${\bf q=0}$, ${\bf k}_{FR}-{\bf k}_{FL}$, $\pm({\bf k}_{FR}+{\bf k}_{FL})$. Note that the contributions with ${\bf q}=\pm 2 {\bf k}_{F}$ vanish by Fermi  statistics.
 
 The long-wavelength expression for the dominant contributions to the spin-nematic correlations, Eq.~(\ref{spin2}), is 
\begin{eqnarray}
\mathcal{F}_{2}({\bf r}) &\sim & -\bigg{[}  \frac{ \left( B_{RR} + B^{*}_{RR}\right)^2}{2\mathfrak{C}_{R} |{\bf r}_c|^3} + R\rightarrow L \bigg{]}~\\
&+&  \frac{\left| B_{RL}+B^{*}_{LR}\right|^2}{\mathfrak{C}^{1/2}_{R}\mathfrak{C}^{1/2}_{L} |{\bf r}_c|^3}\sin[({\bf k}_{FR}-{\bf k}_{FL}) \cdot {\bf r}_c]~\\
&+& \frac{ \left|A_{RL}-A_{LR}\right|^2}{\mathfrak{C}^{1/2}_{R} \mathfrak{C}^{1/2}_{L} |{\bf r}_c|^3} \cos[ ({\bf k}_{FR}+{\bf k}_{FL})\cdot {\bf r}_c] ~,
\end{eqnarray}
where we abbreviate $A^{a=2;a'=4}_{PP'}=A_{PP'}$ and $B^{a=2;a'=4}_{PP'}=B_{PP'}$. The above long-wavelength descriptions can be used to analyze the data obtained by exact numerical calculations.  Here we also note that the model does not have time-reversal and inversion symmetries, so the location of the corresponding R-L patches which are parallel or antiparallel to the observation direction can not be determined easily and need to be found numerically. 

Before leaving this subsection, we remark that we can similarly analyze local energy operators such as bond energy,\cite{Lai11_bondcorr} $\mathcal{B}_{jk} \equiv i u_{jk} \sum_\beta c^\beta_j c^\beta_k$; the long wavelength description contains terms $\sum_\beta f^\beta_P f^\beta_{P'}$ and $\sum_\beta f^{\beta\dagger}_P f^\beta_{P'}$ that are spin-singlet variants of terms in Eq.~(\ref{quadrupolar:dia}).  We can therefore see that the spin, spin-nematic, and local energy observables cover all fermionic bilinears.  Below, we focus on the spin and spin-nematic operators.

\subsection{Exact numerical calculation}\label{Subsec:Data4SU(2)}

\begin{figure}[t]
\subfigure[Spin correlation, Eq.~(\ref{spin0}), for ${\bm r} = x \hat{x}$]{\label{spin_corr_alongx} \includegraphics[width=\columnwidth]{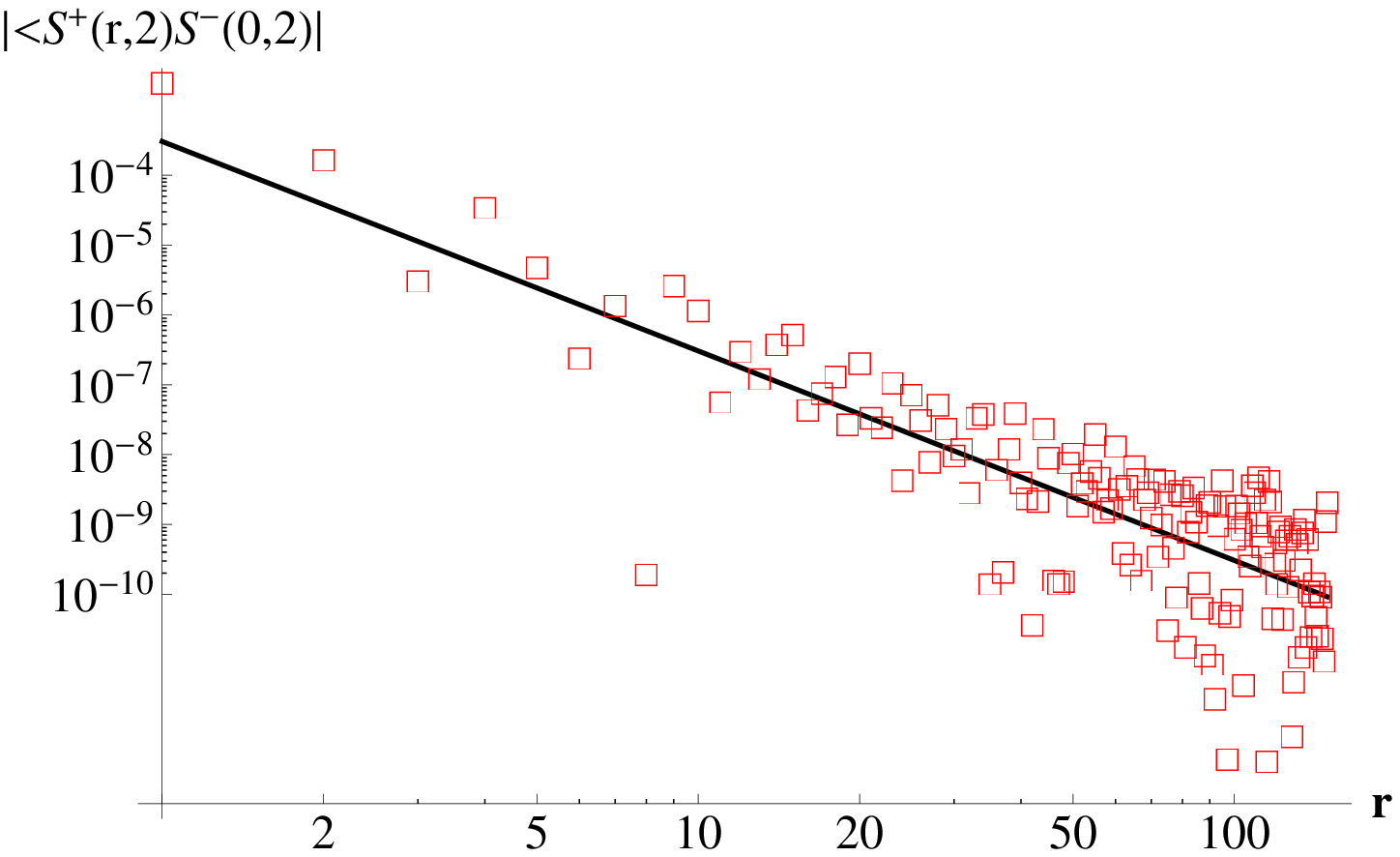}}
\subfigure[Spin-nematic correlation, Eq.~(\ref{spin2}), for ${\bm r} = x \hat{x}$]{\label{nematic_corr_alongx}\includegraphics[width=\columnwidth]{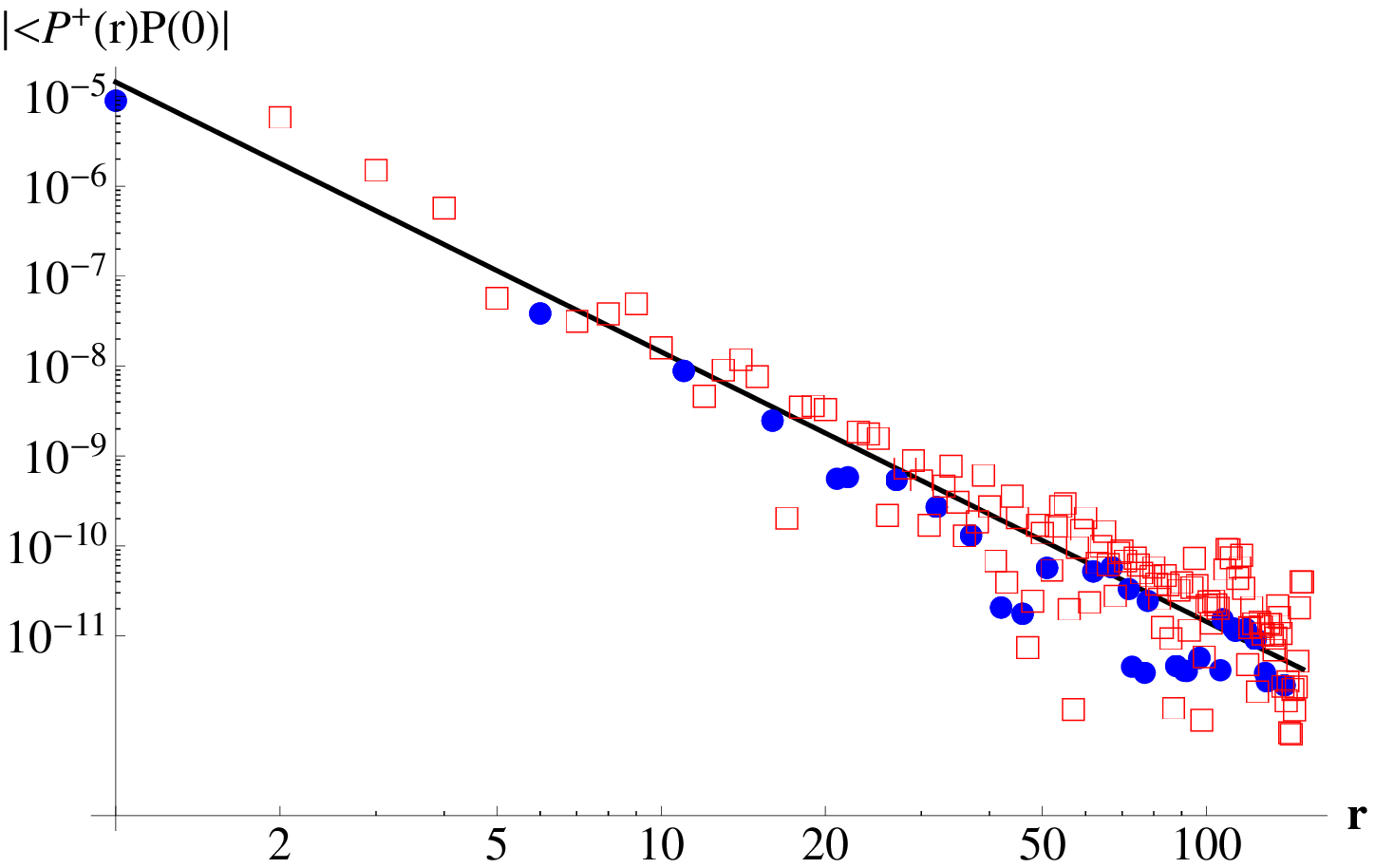}}
\caption{ Figures (a) and (b) illustrate power-law behaviors of the spin correlations and spin-nematic correlations. We calculate $\mathcal{F}_{1}({\bf r})$ and $\mathcal{F}_2({\bf r})$ with ${\bf r}$ taken along the $\hat{x}$-axis for a system containing 300$\times$300 unit cells. The log-log plots in (a) and (b) clearly show the {\it same} dominant $1/x^3$ envelope (straight line in the figures). Here, we show the absolute values of $|\mathcal{F}_{1}({\bf r})|$ and $|\mathcal{F}_{2} ({\bf r})|$, and indicate the sign with open square boxes for negative correlations and filled circles for positive correlations. The irregular behaviors are due to oscillating parts.  The reason that the panel (a) has no positive data is likely because the non-oscillating part is quantitatively stronger, although we can still see oscillations about the $1/x^3$ line.}  
\label{correlators_along_x}
\end{figure}

We calculate the spin correlations, Eq.~(\ref{spin0}), and the spin-nematic correlations, Eq.~(\ref{spin2}), for any real-space separations ${\bf r}$ and confirm that they have the same dominant power law envelope 1/$|{\bf r}|^3$. For an illustration, we show the spin correlations and spin-nematic correlations for ${\bf r}$ along a specific direction, e.g.\ $\hat{x}$-axis, calculated on a 300$\times$300 lattice. In Fig.~\ref{correlators_along_x}, the log-log plot of $|\mathcal{F}_{1} ({\bf r})|$ and $|\mathcal{F}_{2} ({\bf r})| $ clearly shows the same $1/|{\bf r}|^3$ envelope. In addition, the irregular behavior of the data is due to oscillating components. The wavevectors of the real-space oscillations form some singular surfaces in the momentum space, which we analyze next.

\begin{figure}[t]
\subfigure[Three-dimensional view of the spin structure factor]{\label{3dstructure:spin} \includegraphics[width=\columnwidth]{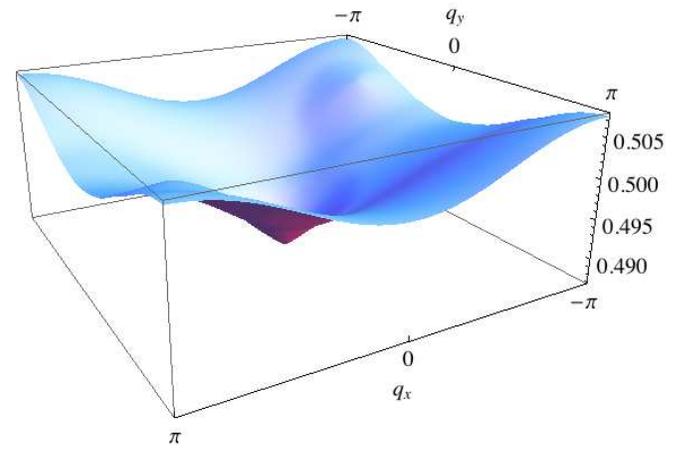}}
\subfigure[Top view of the spin structure factor]{\label{Topview:spin}\includegraphics[width=\columnwidth]{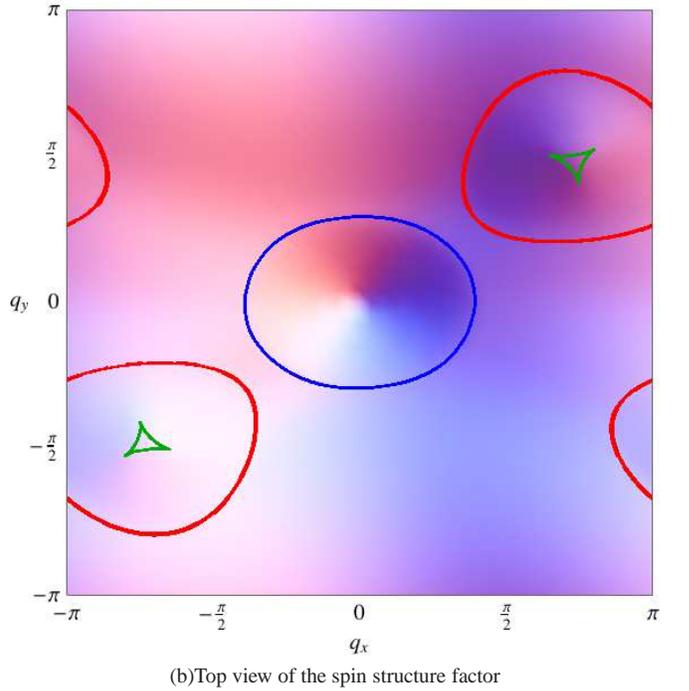}}
\caption{ (a) 3D view of the spin structure factor, $\mathcal{D}_{1}({\bf q})$, defined in Eq.~(\ref{structure_factor}). We can clearly see the singularity $\mathcal{D}_{1} ({\bf q}) \sim |{\bf q}|$ at ${\bf q}=0$ and we also see weak singular lines: one forming central closed ring; two closed rings sitting roughly on one diagonal of the {\bf B.Z.}; and additional weak singular features near the centers of the latter rings. (b) These singular lines are brought out more clearly when the structure factor is viewed from top. We superposed the locations of the singularities calculated using the Fermi surface information: The inner blue ring specifies the line at ${\bf k}_{FR}-{\bf k}_{FL}$; the red closed rings specify the lines at $\pm2{\bf k}_{F}$; the small green triangles specify the lines at $\pm({\bf k}_{FR}+{\bf k}_{FL})$.}  
\label{structure_factor:spin}
\end{figure}

Focusing on the structure factors $\mathcal{D}_{1/2}({\bf q})$ defined in Eq.~(\ref{structure_factor}), we calculate the spin correlation and spin-nematic correlation at each site within a 100$\times$100 lattice and numerically take Fourier transform. Figure~\ref{3dstructure:spin} gives a three-dimensional (3D) view of the spin structure factor.  We can clearly see cone-shaped singularity at ${\bf q}=0$, which is expected from Eq.~(\ref{uniform:spin}):
\begin{eqnarray}
\mathcal{D}_1 ({\bf q\sim 0}) \sim |{\bf q}|.
\end{eqnarray}

A closer look at the spin structure factor also reveals singular surfaces at ${\bf k}_{FR}-{\bf k}_{FL}$, $\pm2 {\bf k}_{F}$, and $\pm({\bf k}_{FR}+{\bf k}_{FL})$, as expected from Eqs.~(\ref{krminuskl:spin})-(\ref{krpluskl:spin}). In order to see the locations of the singular surfaces more clearly and compare with our long-wavelength analysis, we show top view of $\mathcal{D}_1 ({\bf q})$ in Fig.~\ref{Topview:spin}. We numerically calculate the wavectors $2{\bf  k}_{FP}$ and $ {\bf Q}_{\pm} ={\bf k}_{FR}\pm {\bf k}_{FL}$ for all observation directions (by first finding corresponding Right and Left Fermi points with anti-parallel group velocities) and superpose the traced lines on the figure. We see that the lines we get from the long-wavelength analysis match the singular features in the exact spin structure factor. Note that the singularities are expected to be one-sided,
\begin{eqnarray}
&& \mathcal{D}_1 ({\bf Q}_{-} + \delta {\bf q})\sim |\delta q_{||}|^{3/2} \Theta (-\delta q_{||}) ~,\\
&&\mathcal{D}_1 (2 {\bf k}_{FR} +\delta {\bf q})\sim  |\delta q_{||}|^{3/2} \Theta (-\delta q_{||}) ~,\\
&&\mathcal{D}_1({\bf Q}_{+} +\delta {\bf q})\sim |\delta q_{||} |^{3/2} \Theta [-\delta q_{||} {\rm sign}(\mathfrak{C}_R - \mathfrak{C}_L)]. ~~~
\end{eqnarray}
The first and second equations are singular from the inner side of the central ``ring" and the closed rings sitting roughly on one diagonal of the {\bf B.Z.} in Fig.~\ref{Topview:spin}, and the last equation is singular from the inner side of the small ``triangles".

\begin{figure}[t]
\subfigure[Three-dimensional view of the spin-nematic structure factor ]{\label{3dstructure:nematic} \includegraphics[width=\columnwidth]{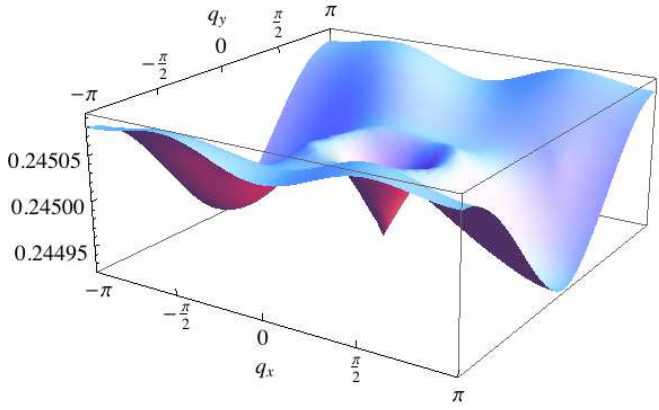}}
\subfigure[Top view of the spin-nematic structure factor]{\label{Topview:nematic}\includegraphics[width=\columnwidth]{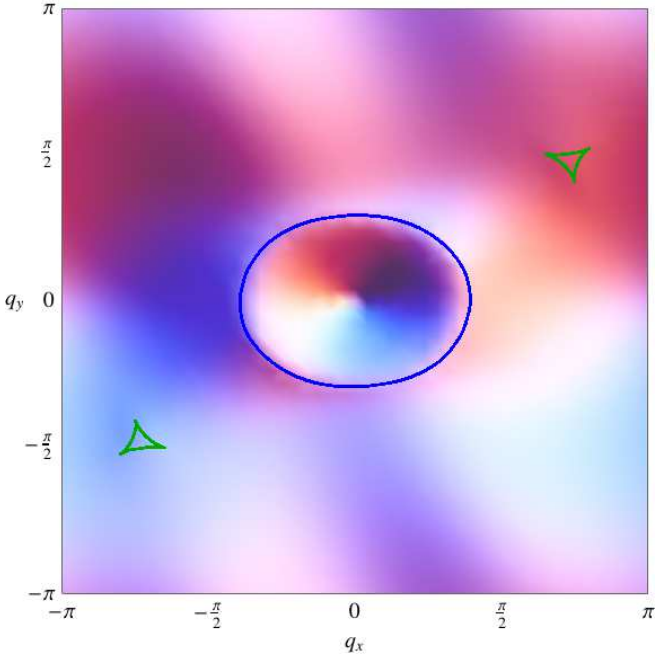}}
\caption{ (a) 3D view of the spin-nematic structure factor, $\mathcal{D}_{2} ({\bf q})$, defined in Eq.~(\ref{structure_factor}). Like the spin structure factor, there is a clear singularity $\mathcal{D}_2 ({\bf q})\sim |{\bf q}| $ at ${\bf q}=0$ and there are weak singular lines, one forming central closed ring and the other forming small triangles sitting roughly on one diagonal of the {\bf B.Z.}. (b) The view from the top shows more clearly the location of the singular lines. We superposed the locations of the singularities calculated using the Fermi surface information: The inner blue ring specifies the line at ${\bf k}_{FR}-{\bf k}_{FL}$, and the small green triangles specify the lines at $\pm({\bf k}_{FR}+{\bf k}_{FL})$. Note that unlike the spin structure factor, there are no $\pm2{\bf k}_{F}$ singularities.}  
\label{structure_factor:nematic}
\end{figure}

Similar analysis can be applied to the spin-nematic structure factor except that there are no $\pm2 {\bf k}_{F}$ singularities. The 3D view of the spin-nematic structure factor is shown in Fig.~\ref{3dstructure:nematic} and we can clearly see the ${\bf q}=0$ singularity and ${\bf q}= {\bf k}_{FR}-{\bf k}_{FL}$ singular line (central ring). The ${\bf q}=\pm ({\bf k}_{FR}+{\bf k}_{FL})$ singular lines (small triangles) are quite weak but still visible, and their locations can be seen more clearly in the view from top shown in Fig.~\ref{Topview:nematic}.  

\section{Majorana spin liquid in the Zeeman field}\label{Sec:Zeeman}

In the presence of the Zeeman magnetic field, we need to consider the additional term in the Hamiltonian
\begin{eqnarray}\label{H_Zeeman}
\mathcal{H}_{Z}=-B_z \sum_{i} \sigma^{z}_{i}= B_z\sum_{{\bf r},a} ic^{x}({\bf r},a)c^y({\bf r},a),
\end{eqnarray}
where we used explicitly the rewriting of spin in terms of Majoranas, Eq.~(\ref{Majorana_reps:spin}).  We remark that the Zeeman magnetic field only couples to the spin degrees of freedom ($c^\alpha$ Majoranas) and not to the orbital degrees of freedom ($d^{\alpha}$ Majoranas that produce the $Z_2$ gauge fields).  Therefore, the model is exactly solvable even in the presence of the magnetic field.  Throughout, we assume the $K_{p}$ terms, Eq.~(\ref{Def:modelH}), are large enough so that the ground state remains in the zero $Z_2$ flux sector.

It is interesting to note that the Zeeman term only affects the $c^x$ and $c^y$ Majoranas while leaving the $c^z$ Majorana unaltered.\cite{Yao10, Biswas11} We can diagonalize the Hamiltonian by starting with the zero-field solution, Eqs.~(\ref{usual_fermion})-(\ref{usual_fermion_H}). We define two complex fermion fields, 
\begin{eqnarray}\label{spin-1 fermion}
 f_{b,+/-}^ \dagger ({\bf k})\equiv [f^{x\dagger}_{b}({\bf k})\pm i f^{y\dagger}_{b}({\bf k})]/\sqrt{2} ~.
\end{eqnarray}
The Hamiltonian in the Zeeman magnetic field becomes
\begin{eqnarray}\label{Hamiltonian: Zeeman}
\mathcal{H} &=& \sum_{b=1}^2 \sum_{{\bf k}\in{\bf B.Z.}} \left( 2\epsilon_{b}({\bf k}) + 2B_z\right)\left[f_{b,-}^\dagger ({\bf k})f_{b,-}({\bf k}) - \frac{1}{2} \right]\hspace{0.5cm}\\
 & + & \sum_{b=1}^2 \sum_{{\bf k}\in {\bf B.Z.}} \left (2\epsilon_{b} ({\bf k}) - 2B_z \right) \left[ f_{b,+}^\dagger ({\bf k})f_{b,+}({\bf k}) - \frac{1}{2} \right]~~\\
 &+& \sum_{b=1}^2  \sum_{{\bf k}\in {\bf B.Z.}} 2\epsilon_{b}({\bf k}) \left[ f^{z\dagger}_{b}({\bf k})f^{z}_{b}({\bf k}) - \frac{1}{2} \right].~~
\end{eqnarray}
This form implies that $f_{b,+}^{\dagger}$ carries $S^z$ quantum number $+1$ and $f_{b,-}^{\dagger}$ carries $S^z=-1$, while $f^{z\dagger}$ carries $S^{z}=0$.

An interesting property in this model is that the $f^z$ Fermi surface (associated with the $c^z$ Majorana) remains no matter how large the magnetic field is;\cite{Biswas11} therefore, there are {\it always} gapless excitations in this system.

\begin{figure}[t]
\includegraphics[width=\columnwidth]{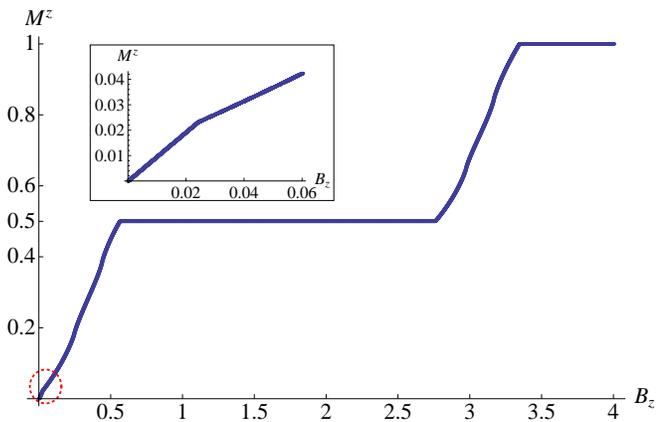}
\caption{The magnetization per site,
$M^z \equiv \la \sum_j \sigma_j^z \ra / N_{\rm sites}$, can be calculated as $(\rho_{1,+} + \rho_{2,+} - \rho_{1,-} - \rho_{2,-})/2$, where $\rho_{b,\pm}$ is the population of the $f_{b,\pm}$ fermions per unit cell. The magnetization curve for our model in the Zeeman field shows rich features (see text) including half- and fully-polarized plateaus.  The inset shows a blow-up of the small $B_z$ region where the Fermi surface $f_{2,-}$ disappears and there is a slope discontinuity. We note that the magnetization inside the unit cell is not completely uniform (because of the reduced lattice symmetries) but is quantitatively similar at each site.}
\label{magnetization}
\end{figure}

For an illustration of several different phases that can occur under the magnetic field, we take the same parameters as in Fig.~\ref{Complex_fermion_bands} and examine the effective Zeeman shifting of the band-1 and band-2 for each complex fermion species. 
Figure~\ref{magnetization} shows the magnetization as a function of magnetic field $B_z$. There are quite rich features in this model. When we turn on the magnetic field, the $f_{b,+}$ bands move downwards while the $f_{b,-}$ bands move upwards. First, the field increases up to a threshold value, roughly $B_z = 0.024$, where the Fermi surface of the $f_{2,-}$ vanishes and there is a discontinuity in the slope of the magnetization curve shown in the inset in Fig.~\ref{magnetization}. The $f_{2,+}$ band is pushed down and the $f_{2,+}$ Fermi sea keeps growing until it completely covers the {\bf B.Z.} at $B_z \simeq 0.5$.  For the field between 0.5 and 2.8, the $f_{1,+}$ band remains above the zero energy, and we have the half-polarized magnetization plateau phase. The $f_{1,+}$ band reaches zero energy at $B_z \simeq 2.8$, we leave the first plateau phase and the magnetization starts to increase.  When the magnetic field is large enough to completely push the $f_{1,+}$ band below zero, $B_z \simeq 3.3$, the $f_{1,+}$ Fermi surface also vanishes and we enter the fully-polarized phase, the second plateau phase.  We also see some weak features in the regimes of increasing magnetization that are due to the van Hove singularities when the energy passes the saddle points of bands 1 or 2 [cf.\ the contour plot of band 2 in Fig.~\ref{fermi_surface}], but these van Hove singularities are rather weak in 2D.

We remark that in all regimes, the Fermi surface of the $f^z_{2}$ remains the same and gives gapless excitations. The fully polarized phase is  actually the original Kitaev-type model proposed by Baskaran\etal .\cite{Baskaran09}  This can be seen either directly by examining the physical Hamiltonian Eq.~(\ref{Def:modelH}), or in the Majorana representation where $\sigma^z_j = 1 = -i c^x_j c^y_j$, so the constraint becomes $D_j = -i c^x_j c^y_j c^z_j d^x_j d^y_j d^z_j =c^z_j d^x_j d^y_j d^z_j=1$, and the model in terms of the orbital degrees of freedom reduces to that in Ref.~[\onlinecite{Baskaran09}].

It is interesting that in the half-polarized plateau phase, even though the spin excitations are gapped, the spin degrees of freedom are entangled in the ground state.  The spinless gapless excitations can be in principle detected by measuring bond-energy correlations or by entanglement entropy calculations.
Finally, the regimes of increasing magnetization can be viewed as generic compressible Bose-metals\cite{DBL} in the model with global U(1) symmetry.

\section{Discussion}\label{Sec:discussion}

We proposed SU(2)-invariant Kitaev-type model on the decorated square lattice that realizes QSL with parton Fermi surfaces.  Having the benefit of the exact solutions, we can make robust general observations about such SU(2)-invariant Majorana QSL.  One of the distinguishing characteristics of this state is that it has strong spin and spin-nematic fluctuations as manifested by the same power-law behavior in the correlations considered in our work.  Because of the finite density of states at the Fermi surface, properties such as the specific heat, spin susceptibility, and NMR relaxation rates are essentially similar to a metal.  Note that the volume/shape of the Fermi sea can be arbitrary, and we can easily tune the model to have larger or smaller Fermi pockets.  In this way, the phase is quite distinct from the conventional spinon Fermi sea QSL,\cite{LeeNagaosaWen} where half of the Brilloin zone is populated by spin-1/2 spinons before the Gutzwiller projection.  Given the variability of the Fermi surface, the properties of the Majorana QSL can be very sensitive to parameters, which we can also easily tune to produce gapped phases.

Let us briefly discuss stability of the exactly solvable model to general perturbations. First, we note that our complex fermions $f_\alpha$ are not conserved microscopically.  In principle, allowed four-fermion interactions would contain terms such as $f^\dagger_\alpha f^\dagger_\beta f^\dagger_\gamma f^\dagger_\delta$ and $f^\dagger_\alpha f^\dagger_\beta f^\dagger_\gamma f_\delta$, in addition to the more familiar terms $f^\dagger_\alpha f^\dagger_\beta f_\gamma f_\delta$.   The origin of the non-conservation of the complex fermions is because the microscopic fields are Majorana fermions, while we used the complex fermions as a convenient tool to reduce the problem to more familiar calculations.  However, if the symmetries of the model are sufficiently low, there can be an ``emergent" conservation of $f$-s.  For example, if in our treatment we have a very small Fermi pocket of $f$-s centered at some momentum ${\bf K} = (K_x, K_y)$, then the four-fermion terms $f^\dagger f^\dagger f^\dagger f^\dagger $ carry approximate momentum $4 {\bf K}$ while $f^\dagger f^\dagger f^\dagger f$ carry $\approx \! 2 {\bf K}$, so these terms are not allowed by momentum conservation if ${\bf K}$ is some generic non-special wavevector.  In this case, only the familiar terms $f^\dagger f^\dagger f f$ conserving the fermion number are left and we have the emergent fermion number conservation law in the long-wavelength theory.   The above is also true in our paper in which the non-symmetric small Fermi pockets are realized.  Furthermore, only benign forward-scattering four-fermion terms survive in our model since the Cooper-pair interactions carry non-zero momentum.  We do not need to consider six- and eight-fermion terms as they are irrelevant in the Renormalization Group(RG) sense.

We have chosen a model that lacks time-reversal and lattice point group symmetries so as not to worry about possible residual pairing instabilities away from the exactly solvable limit.  Such instabilities can be relatively weak also in more symmetric models and the discussed phenomenology can apply in these cases as well. It should be possible to explore the stability and the nearby phases by studying such models on ladders\cite{SBM_Solvay} using weak-coupling RG technique and Bosonization analysis.\cite{Balents96,Lin97,Shankar_Acta,Shankar_RGRMP}

In the presence of the Zeeman magnetic field, there are more interesting phases with distinct stable Fermi pockets of $f_{b,+}$, $f_{b,-}$, and $f_b^z$. The Zeeman field breaks the global SU(2) down to U(1), and the compressible phases in the field are {\it Bose-Metal}-like phases.\cite{DBL, Sheng09}  

We also found an interesting plateau phase at half-magnetization. Due to the gap for spin excitations, it is a spin insulator, but since the $f^z$ remains gapless, we expect to still have metal-like specific heat and thermal conductivity. 

This behavior in the plateau phase arises because some of the parton constitutents of the spin operator acquire a gap (band gap in the present case).  Some such physics perhaps can be relevant for the explanation of very recent NMR experiments\cite{Itou10} in \dmit\ showing a drastic reduction in the spin relaxation below temperature of the order 1K as if a spin gap opens up, while the thermal conductivity measurements and thermodynamic measurements\cite{MYamashita10, SYamashita11} are consistent with the presence of a Fermi surface of fermionic excitations down to the lowest temperatures. This phenomenology is also qualitatively similar to our recent paper\cite{SBMZeeman} working in a setting closer to the \dmit~experiments.  We considered a scenario in which, upon writing the spin operator as $S^{+} = f^{+}_{\uparrow} f_{\downarrow}$, there could be a phase in which one spinon species becomes gapped due to pairing, while the other species retains the Fermi surface.  It is fascinating to further explore such idea where some partons are gapped and some are gapless in more realistic settings.

\acknowledgments
This research is supported by the National Science Foundation through grant DMR-0907145 and by the A.~P.~Sloan Foundation.

\bibliography{biblio4MajoranaSL}

\begin{thebibliography}{71}
\bibitem[{foo()}]{footnote_edge}
\bibinfo{note}{Furthermore, we also calculated the energy spectrum in a
  cylindrical geometry with one open boundary and confirmed that there are no
  gapless edge modes in this system}.
\end{thebibliography}
\end{document}